\documentclass{camera}
\usepackage{graphicx}  

\begin{document}


\title{Nuclear Reactions at Intermediate Energies{\footnote{Plenary
talk presented at NN2015, June 21-26, 2015, Catania, Italy}}}

%
\author{Radhey Shyam}

%
\organization{Saha Institute of Nuclear Physics, 1/AF Bidhan Nagar, 
Kolkata 700064, India}

\maketitle

\begin{abstract}
In the domain of Nuclear reactions at intermediate energies, the QCD 
coupling constant $\alpha_s$ is large enough ($\sim$ 0.3 - 0.5) to render 
the perturbative calculational techniques inapplicable. In this regime 
the quarks are confined into colorless hadrons and it is expected that
effective field theories of hadron interactions via exchange of hadrons, 
provide useful tools to describe such reactions. In this contribution 
we discuss applications of one such theory, the effective Lagrangian 
model, in describing the hadronic reactions at intermediate energies 
whose measurements are the focus of a vast international experimental 
program.
\end{abstract}
In the quest for understanding the quantum chromodynamics (QCD), the 
hadron-hadron, hadron-photon and hadron-lepton reactions have played a 
major role. From these studies two quite distinct regimes of QCD have
emerged. The first regime is of those processes that involve very high 
momentum transfers.  Here the asymptotic freedom property of QCD has 
enabled the perturbative QCD (pQCD) methods to describe~\cite{bet07} a 
large set of high energy large momentum transfer reaction observables 
with great precision, providing the most precise test of QCD to date
\cite{col13,mul81}.

The second regime that includes nuclear reactions at intermediate energies,
is characterized by the confinement of quarks inside the hadrons. This 
remarkable feature, along with the running of the QCD coupling constant
($\alpha_s$) towards larger values makes the pQCD methods inapplicable - 
one requires techniques that can solve the theory exactly. To date, the 
only fully predictive non-perturbative method for studying QCD at low 
energies is lattice QCD (LQCD). The progress in LQCD studies of the 
low-lying spectrum of QCD, hadronic structure and interactions has been 
significant in the recent years (see, eg., Ref.~\cite{bri14}). 

Nevertheless, LQCD calculations are still far from being routinely amiable 
to solutions of intermediate energy nuclear reactions. It is therefore, 
necessary to use effective methods to describe the dynamics of such 
processes. The effective field theories where hadrons interact via exchange 
of hadrons, provide a viable alternative to LQCD for the description of the 
intermediate energy nuclear reactions (see, e.g., Refs.
\cite{nie10,clo14,dev13,sie14,feu98,pen02,shy08}). These theories implement 
the symmetries of the underlying theory, QCD, by using proper effective 
Lagrangians. However, many of these models invariably require {\it a priory} 
knowledge of unknown parameters related to the coupling constants associated 
with the effective Lagrangians and the masses of the hadrons. Because the 
effective theories provide flexible theoretical approaches that allow a 
swift understanding of the intricacies involved in the scattering problems 
at intermediate energies and have the ability to guide to further experimental 
studies, steady progress has been made towards determining the realistic input 
parameters involved in these theories.

In this presentation we discuss applications of effective Lagrangian based 
theoretical models, to describe the meson production reactions (both non-strange 
and strange) in photon and meson induced reactions on nucleons at intermediate 
energies. We also present some results on the charmed baryon and meson production 
in the antiproton-proton annihilation process.   

\section{Meson production in photon-nucleon reactions}

The photon induced reactions on nucleons where mesons and baryons are measured 
as final decay products, have revealed a rich excitation spectrum of nucleons 
and other baryons that reflects their complicated multi-quark inner dynamics. 
The determination of the properties of the nucleon excited states (or resonances) 
from the data of such experiments is a major challenge of hadron physics 
\cite{cap00}.  

The effective methods that are used to describe the dynamics of meson production 
reactions, include explicitly the baryon resonance states whose properties are 
extracted by comparing the predictions of the theory with the experimental data
\cite{feu98,pen02,shy08,sag01,bor02,jul08,shy10,ron15}. However, for a reliable 
realization of this task, a model is required that can analyze different 
reactions over the entire energy range using a single Lagrangian density, which 
generates all non-resonance contributions from Born, $u$- and $t$-channel 
contributions without introducing new parameters. At the same time, the Lagrangian 
should also satisfy the symmetries of the fundamental theory (i.e. QCD) while 
retaining only mesons and baryons as effective degrees of freedom. The 
coupled-channels method within the $K$-matrix approximation is one model that 
respects these constraints while providing a way to analyze simultaneously all the 
reaction data for a multitude of observables in different reaction channels. It 
also leads to a convenient way of imposing the unitarity constraint because the $S$ 
matrix in this approach is unitary provided the $K$-matrix is taken to be real and 
Hermitian (see, e.g., Ref.~\cite{shy08,shy10}). 

The experimental data on the $\eta$ meson and associated strangeness ($K^+\Lambda$ 
and $K^+\Sigma^0$) production in photon-nucleon reactions are available for photon 
energies ($E_\gamma$) in the range of threshold to 3~GeV. Thus they cover not only the 
entire resonance region but also the region where the background contributions 
($t$-channel amplitudes mainly) are expected to be dominant. The data for the $\eta$ 
channel are reviewed recently in Ref.~\cite{kas15}. The coupled-channels $K$-matrix 
method has been used extensively in Refs.~\cite{shy08,shy10} to describe the data of 
both these channels in the entire range of $E_\gamma$ with a single set of parameters. 
These reaction channels have also been studied within the Giessen coupled-channels 
$K$-matrix model in Refs.~\cite{shk13,cao13} for $E_\gamma$ between threshold 
to 2.0 GeV. 

\begin{figure}[t]
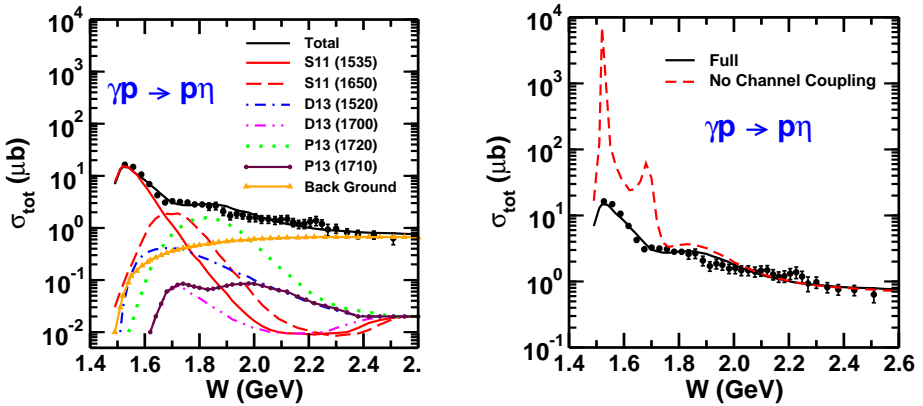

\begin{tabular}{cc}
\includegraphics[width=0.45 \textwidth]{Fig01a.eps} & \hspace{0.10cm}
\includegraphics[width=0.45 \textwidth]{Fig01b.eps}
\end{tabular}
\caption{({\bf Left panel}) Total cross section for the $\gamma p \to \eta p$ 
reaction as a function of $\gamma p$ invariant mass $W$. Individual contributions 
of various resonances  and of background terms are shown by different curves as 
indicated in the figure. ({\bf Right panel}) Effect of channel coupling on the total 
cross section for the $\gamma p \to \eta p$ reaction as a function of $W$. The 
solid line represents the results of the full coupled-channel calculations  
while the dashed line shows the one where channel coupling is switched off.
}
\label{Fig01} 
\end{figure}
\noindent

As an example of the success of the coupled-channels $K$-matrix model of Refs.
\cite{shy08,shy10} in describing the meson production in photon-nucleon reactions, we 
show in the left panel of Fig.~1 the contribution of various resonances to the total 
cross section ($\sigma_{tot}$) of the $\gamma p \to \eta p$ reaction. The experimental 
data are taken from Ref.~\cite{cre05}. We see that the general features of the data are 
described reasonably well by the calculations. While contributions of the $S_{11}$(1535) 
resonance dominate the $\sigma_{tot}$ from near threshold to invariant mass ($W$) values 
of 1.7~GeV (corresponding to E$_\gamma \sim$ 1.1~GeV), those of the $S_{11}$(1650) and 
$P_{13}$ (1720) resonance are important for E$_\gamma$ between 0.950~GeV to 1.28~GeV and 
1.1~GeV to 2.2~GeV, respectively. The  contributions of other resonances are quite weak 
in the entire range of $E_\gamma$. The cross sections beyond 2~GeV are mostly 
governed by the contributions of the background terms.  

The effect of channel coupling on the $\sigma_{tot}$ is shown in the right panel of 
Fig.~1. In the no coupling case (NCC), the amplitudes of various processes are simply 
added together, ignoring the modifications to the widths of the resonances introduced 
by the channel couplings. We notice that for $W > 1.8$~GeV the differences between 
the full and the NCC results are very small because the contributions of the resonances 
are almost negligible in this region. However, at lower energies, the channel-coupling 
effects are large. In fact, at some energies, the resonance propagators can develop poles 
in the absence of channel couplings. Thus modifications introduced to the widths of the 
resonances due to channel couplings are indeed vital for reproducing the energy dependence 
of the experimental cross sections. Therefore, the  use of a full coupled-channels approach 
is essential to describe the meson production in photon-nucleon reactions.  

\section{Double strangeness ($S$ = -2) production with $K^-$ induced reactions}

The investigations of the cascade ($\Xi$) hypernuclei (having strangeness $S$ = -2) are
of great importance due to several reasons. The binding energies and widths of the $\Xi$ 
hypernuclear states determine the strengths of the $\Xi N$ and $\Xi N \to \Lambda \Lambda$ 
interactions, which are difficult to explore directly in a laboratory. This input is vital 
for revealing the entire picture of strong interactions among octet baryons. Since strange 
quarks are negatively charged they are preferred in charge neutral dense matter. Thus these 
studies are capable of probing the role of strangeness in the equation of state at high 
density, eg. in the cores of neutron stars~\cite{bie10}.

Because the $(K^-,K^+)$ reaction leads to the transfer of two units of both charge and 
strangeness to the target nucleus, it provides one of the most promising ways of studying 
the $S = -2$ systems such as $\Xi$ hypernuclei and a dibaryonic resonance ($H$), which is a 
near stable six-quark state with spin parity of $0^+$ and isospin 0~\cite{jaf77,mul83,shy13}. 
The $(K^-,K^+)$ reaction implants a $\Xi$ hyperon in the nucleus through the elementary 
$p(K^-, K^+)\Xi^-$ process. The cross sections for this reaction were measured in the 1960s 
and early 1970s using hydrogen bubble chambers.
\begin{figure}[t]
\begin{center}
\includegraphics[width=0.75 \textwidth]{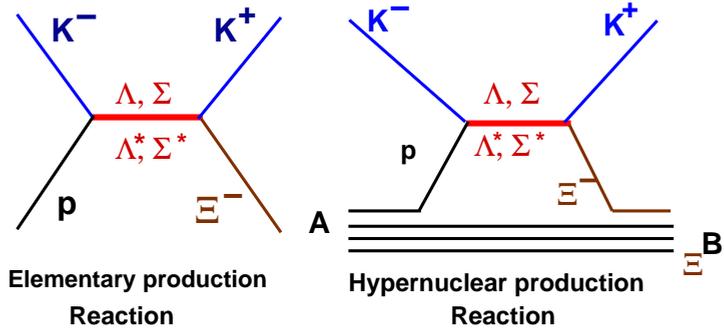} 
\caption{({\bf Left panel}) Graphical representation of the $s$-channel diagrams used to 
describe the $p(K^-, K^+)\Xi^-$ reaction. The $u$-channel diagrams are not explicitly 
shown here but their contributions are included in the calculations. ({\bf Right panel}) 
Diagram used to calculate the amplitudes of the $A(K^-, K^+){_{\Xi^-}}\!\! B$ reaction. 
$A$ represents the target nucleus and ${_{\Xi^-}}\!\! B$ the final hypernucleus.
}
\end{center}
\label{Fig02}
\end{figure}
\noindent

In Refs.~\cite{shy11,jac15}, the  $(K^-,K^+)$ reaction on a proton target leading to 
the production of a free $\Xi^-$ hyperon has been studies within a single-channel 
effective Lagrangian model (SCELM) (see, eg., Refs.~\cite{shy99,shy03}). A 
coupled-channels study of this reaction is not feasible at this stage due to the 
scarcity of the data on the $K^-$ interaction channels. In the SCELM the full field 
theoretic structure of the interaction vertices are retained and baryons are 
treated as Dirac particles. The initial state interaction of the incoming $K^-$ with 
a free or bound target proton leads to the excitation of intermediate $\Lambda$ and 
$\Sigma$ resonant states, which propagate and subsequently decay into $\Xi^-$ and $K^+$. 
In case of the reaction on nuclei, $\Xi^-$  gets captured into one of the nuclear 
orbits, while the $K^+$ meson goes out. We have considered as intermediate states the 
$\Lambda$ and $\Sigma$ hyperons and eight of their resonances with masses up to 2.0 GeV, 
which are represented by $\Lambda^*$ and $\Sigma^*$ in Figs.~2. The effective Lagrangians 
and the coupling constants at various vertices are taken to be the same as those described 
in Ref.~\cite{shy11}.
\begin{figure}[t]
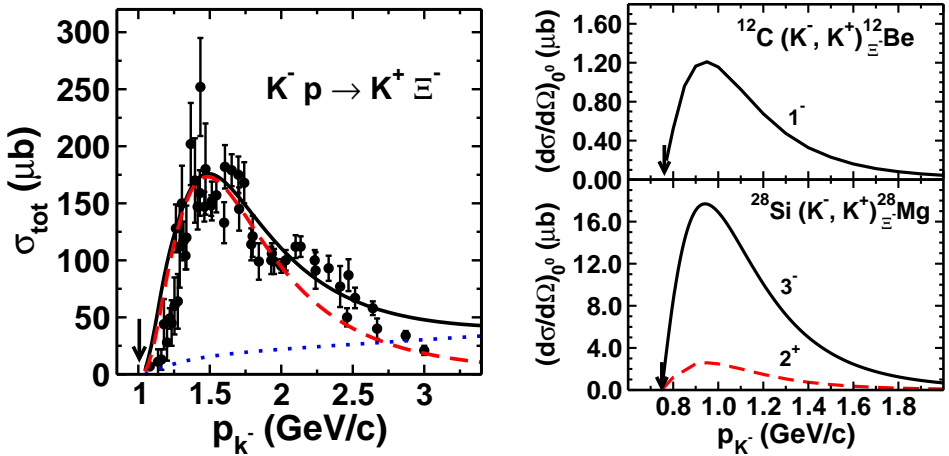

\begin{tabular}{cc}
\includegraphics[width=0.50 \textwidth]{Fig03a.eps} & \hspace{0.10cm}
\includegraphics[width=0.43 \textwidth]{Fig03b.eps}
\end{tabular}
\caption{({\bf Left panel}) Total cross section for the $p(K^-, K^+)\Xi^-$ reaction as a
function of incident $K^-$ momentum. Individual contributions of of $s$- and $u$-channel 
diagrams to the total cross sections are also shown. ({\bf Right panel}) $p_{K^-}$ dependence 
of the zero degree differential cross section for the $^{12}$C$(K^-,K^+)^{12}{\!\!\!_{\Xi^-}}
$Be and $^{28}$Si$(K^-,K^+)^{28}{\!\!\!_{\Xi^-}}$Mg reactions. The spin-parity of the final 
hypernuclear states are indicated on each curve. Arrows show the thresholds for the 
corresponding reactions. 
}
\label{Fig03} 
\end{figure}
\noindent
 
In the left panel of Fig.~3, we show the comparison of the SCELM calculations with the data 
for the $\sigma_{tot}$ of the $p(K^-, K^+)\Xi^-$ reaction as a function of $K^-$ beam 
momentum ($p_{K^-}$). It is clear that the data are reproduced reasonably well by the 
model within the statistical errors. The zero degree differential cross sections for 
the hypernuclear production reactions $^{12}$C$(K^-, K^+)^{12}{\!\!\!_{\Xi^-}}$Be and 
$^{28}$Si$(K^-,K^+)^{28}{\!\!\!_{\Xi^-}} $Mg calculated by using the same vertex 
parameters, are shown in the right panel of Fig.~3. In these calculations we have 
employed pure single-particle-single-hole $(p^{-1}\Xi)$ configurations to describe the 
nuclear structure part. The bound proton-hole and $\Xi^-$-particle state spinors were 
generated in a phenomenological model, where they are obtained by solving the Dirac 
equation with scalar and vector fields having Woods-Saxon radial forms.  With a set of 
radius and diffuseness parameters, the depths of these fields are searched to reproduce 
the binding energy (BE) of the given state. The BEs for the $\Xi^-$ bound states were taken 
to be those predicted in the latest version of the quark-meson coupling (QMC) model
\cite{kaz12,shy13a} as their experimental values are still unknown.. In this figure, we 
have shown results for populating the hypernuclear state with maximum spin of natural parity 
for each configuration. We see that the zero degree differential cross sections for the 
$\Xi^-$ hypernuclear production reactions peak around the beam momentum of 1.0 GeV/c with 
peak cross section of more than 1 $\mu b$. They closely follow the trends of the elementary 
$\Xi^-$ production cross sections. Our predictions will be useful for the future JPARC 
experiments.
 
\section{Charmed hadron production in ${\bar p}p$ annihilation }

The heavy flavored hadrons consisting of relatively heavier mass charm quarks 
provide an additional handle for the understanding of QCD. The charm quark introduces 
a mass scale that is much larger than the confinement scale $\Lambda_Q \approx 300$ 
MeV. In contrast, the energy scale of the lighter quarks is $\ll \Lambda_Q$.  The 
presence of two scales in such systems naturally leads to the construction of an effective 
theory where one can actually calculate a big portion of the relevant physics using 
perturbation theory and renormalization-group techniques.
 
In this context, the investigations of the production of heavy flavor hadrons
are of great interest. Since the discovery of $J/\psi$ in 1974~\cite{aub74,aug74},
the production of charmonium ($c{\bar c}$) states has been extensively studied
in electron-positron and proton-antiproton (${\bar p}p$) annihilation experiments. 
Yet a substantial part of the charmonium spectrum is still to be precisely measured.
The first charmed baryon states were detected in 1975 in neutrino interactions
~\cite{cas75}. Although many new excited charmed baryon states have been discovered
since then, the studies of the production and spectroscopy of the charmed baryons 
have not been carried out in the same detail as the charmonium states.
 
In the near future, charmed hadron production will be investigated in the  $p{\bar p}$ 
annihilation in the (${\bar P}ANDA$) experiment at the FAIR in GSI, Darmstadt 
\cite{wie11}. The advantage of using antiprotons in the study of the charmed baryon 
is that in $p{\bar p}$ collisions the production of no extra particle is needed for 
the charm conservation, which reduces the threshold energy. Moreover, all the the 
charmonium states can be reached directly in this method, which is in contrast to the  
electron-positron annihilation where direct formation is allowed for only those final 
charmonium states that have the quantum numbers of the photon ($J^{PC} = 1^{--}$). 

For the planning of these experiments at the ${\bar P}ANDA$ facility, reliable 
theoretical estimates of the cross section of ${\bar p} p \to {\bar \Lambda}_c^- 
\Lambda_c^+$, ${\bar p} p \to {\bar D}^0 D^0$ and ${\bar p} p \to D^- D^+$ reactions   
are of crucial importance~\cite{hai10,hai14}. In Ref.~\cite{shy14}, the SCELM has 
been used to study the ${\bar p} p \to {\bar \Lambda}_c^- \Lambda_c^+$ reaction as a 
sum of the $t$-channel $D^0$ and $D^{*0}$ meson-exchange diagrams [Fig.~4(a)]. A 
similar model is employed in Ref.~\cite{shy15} to calculate the  cross sections of 
${\bar p} p \to {\bar D}^0 D^0$, and $ {\bar p} p \to D^- D^+$ reactions as a sum 
of the $t$-channel $\Lambda_c^+$ and $\Sigma_c^+$, and $\Sigma_c^{++}$ baryon-exchange 
diagrams, respectively [Figs.~4(b) and 4(c)]. Also included in these calculations are 
the $s$-channel excitation, propagation and decay of the $\Psi(3770)$ resonance into 
the ${\bar D}^0 D^0$ and $D^- D^+$ channels.  
 
It is found that the $\sigma_{tot}$ of the ${\bar p} p \to {\bar \Lambda}_c^-
\Lambda_c^+$ and ${\bar p} + p \to {\bar D}^0 D^0$ reactions at the ${\bar p}$ beam 
momentum of 15 $GeV/c$ (which the regime of the ${\bar P}ANDA$ experiment) are  
around 25 $\mu$b and 550 $n$b, respectively. These values are drastically larger than 
those predicted in previous calculations based on models that invoke the quark degrees 
of freedom. The future ${\bar P}ANDA$ experiment will clarify the adequacy of various 
models at these relatively lower beam momenta. 
\begin{figure}[t]
\includegraphics[width=0.90 \textwidth]{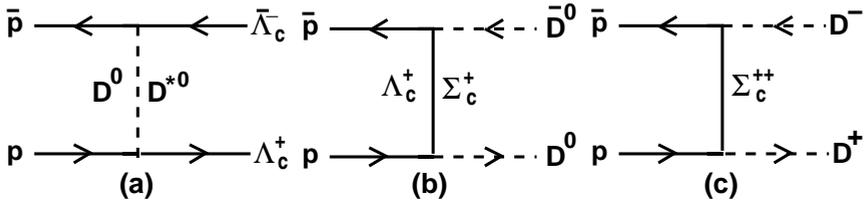} 
\caption{ Graphical representations of the $t$-channel meson exchange diagrams used 
to describe the ${\bar p} + p \to {\bar \Lambda}_c^- + \Lambda_c^+$ reaction
({\bf a}), and the $t$-channel baryon exchange diagrams to describe the ${\bar p} + p 
\to {\bar D}^0 + D^0$ and ${\bar p} + p \to D^- + D^+$ reactions ({\bf b and c}). 
}
\label{Fig04} 
\end{figure}
\noindent


In {\bf summary}, we can say that effective Lagrangian models are able to describe 
reasonably well the intermediate energy nuclear reactions. The coupled-channels 
$K$-matrix method allows for the simultaneous calculation of the observables for a large 
multitude of reactions with considerably fewer parameters than would be necessary if 
each reaction channel were fitted separately. For the strangeness and charm production 
reactions, a single-channel effective Lagrangian method  with suitably chosen 
parameters is able to make important predictions that will be tested in the future 
experiments planned at upcoming facilities.

{\bf Acknowledgments}: This work has been supported by the Helmholtz International 
Center (HIC) for FAIR and the Council of Scientific and Industrial Research (CSIR), 
India. The author acknowledges useful discussions with H. Lenske, O. Scholten, 
A.W. Thomas and K. Tsushima.

%
\end{document}